\documentclass[aps,prb,twocolumn,unsortedaddress,showpacs]{revtex4-1}
\usepackage{graphicx}
\usepackage{calc}
\usepackage{bm}

\begin{document}

\title{Energy spectrum reconstruction at the edge of a two-dimensional electron system with strong spin-orbit coupling}

\author{A.~Kononov}
\affiliation{Institute of Solid State Physics RAS, 142432 Chernogolovka, Russia}

\author{G.~Biasiol}
\affiliation{IOM CNR, Laboratorio TASC, 34149 Trieste, Italy}

\author{L.~Sorba}
\affiliation{NEST, Istituto Nanoscienze-CNR and Scuola Normale Superiore, 56127 Pisa, Italy}

\author{E.V.~Deviatov}
\email[E-mail: ]{dev@issp.ac.ru}
\affiliation{Institute of Solid State Physics RAS, 142432 Chernogolovka, Russia}

\date{\today}

\begin{abstract}
We experimentally demonstrate the simultaneous reconstruction of the  bulk and the  edge energy spectra  caused by strong spin-orbit interaction, at the two lowest filling factors $\nu=1$ and $\nu=2$ for $In_xGa_{1-x}As$ two-dimensional electron system with $x = 0.75$. The observed reconstruction is of different character at these filling factors: it is characterized by zeroing of the energy gap at the levels crossing point at $\nu=2$, while the reconstruction at $\nu=1$ goes through the coexistence of two $\nu=1$ quantum Hall phases with different spin projections. An analysis indicates a strong influence of many-body interaction on the reconstruction at $\nu=1$.
\end{abstract}

\pacs{73.40.Qv  71.30.+h}

\maketitle

\section{Introduction}

Long-term interest to the investigation of semiconductor systems with strong spin-orbit (SO) coupling is connected, on the one hand, with their intriguing physical properties, and, on the other hand, with possible spintronic applications. For the applications, SO interaction opens a way to electrically control the electron spin.~\cite{datta} For the pure science, SO interaction is an example of the relativistic effect being observable in conventional semiconductors physics.~\cite{winkler} Nowdays,  SO-like Hamiltonian term is essential in theoretical models of systems with nontrivial topology (topological insulators). Thus, the systems with strong SO coupling can help in transferring the well developed methods of semiconductor physics to fields of new  physical concepts. 

From a number of different two-dimensional  systems with strong SO coupling there is one which is characterized by a unique set of parameters. Namely, a two dimensional electron gas (2DEG), confined in a narrow asymmetric $In_xGa_{1-x}As$ quantum well with $x \geq 0.75$, is characterized by (i) high mobility,~\cite{biasiol08} because the well is nominally undoped; (ii) low electron concentration, which can be widely controlled by an external gate~\cite{biasiol08}; (iii) strong Rashba-type SO coupling~\cite{nitta}; (iv) high $g$-factor, subject to exchange-interaction enhancement~\cite{kita,gui,biasiol04} $g^*\leq30$.

This parameter set makes this system very suitable for investigations in the quantum Hall (QH) regime. The transitions between QH states with different spin polarization were investigated in Ref.~\onlinecite{biasiol04} at high filling factors $\nu=4,6,8,10,12$. The transitions were induced by increasing the ratio between the Zeeman and the cyclotron energies in a tilted-field configuration. The obtained $g$-factor was demonstrated to be subject  to   many-body effects~\cite{biasiol04}.

However, even in normal fields, strong SO coupling can cause  transitions between different QH states at low filling factors~\cite{spain}. A single-particle spectrum was originally calculated in Ref.~\onlinecite{bychkov} for Rashba-type SO two-dimensional systems in quantizing magnetic fields. The spectrum is represented by distinct energy levels
\begin{equation}
E_{N_L}^s=\hbar\omega_c\left [N_L+\frac{1}{2}s\sqrt{\left (1-|g|\frac{m}{2m_0}\right )^2+\frac{\gamma}{B}N_L}\right ],
\label{rashba}
\end{equation}
where $s=\pm1$ for $N_L=1,2,3,....$ and $s=+1$ for $N_L=0$, $m=0.035 m_0$ is the effective electron mass~\cite{biasiol04}, $\gamma$ defines the SO coupling strength. 

\begin{figure}
\includegraphics*[width=\columnwidth]{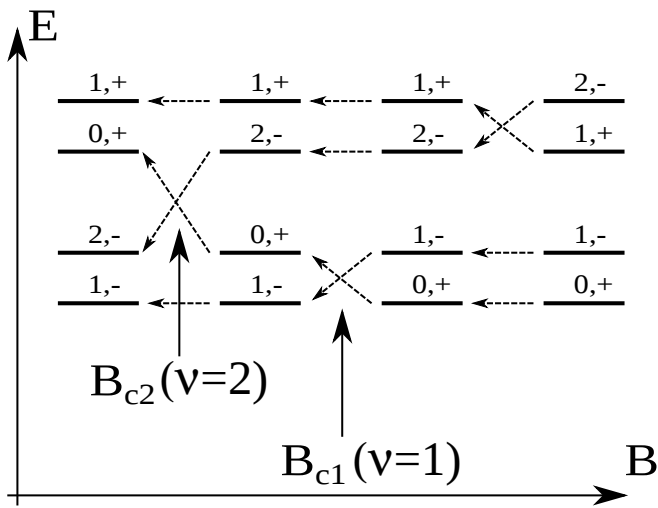}%
\caption{Schematic diagram of the energy levels systematics of Eq.~(\protect\ref{rashba}) for  $N_L=0,1,2$, $s=\pm$.  Only relative positions of the energy levels are shown, disregarding their absolute energy values. In the limit of extremely high fields, energy levels represent the well-known ladder of the spin-split Landau sublevels.  Decreasing of the field $B$ is equivalent to adiabatic switching on the spin-orbit interaction, because of the $\frac{\gamma}{B}$ term  in  Eq.~(\protect\ref{rashba}). This results in the intersections of the levels with different $s$ at critical fields $B_c$, as can be directly obtained from Eq.~(\protect\ref{rashba}).
\label{levels}}
\end{figure}

In the limit of extremely high magnetic fields, the spectrum (\ref{rashba}) represents the well-known ladder of the spin-split Landau sublevels. Diminishing of the magnetic field is equivalent to adiabatic switching on the SO-interaction, because of increasing term $\frac{\gamma}{B}$ in~(\protect\ref{rashba}). It is important, that even  the level {\em systematics} is affected by increasing the SO  term $\frac{\gamma}{B}$, see Fig.~\ref{levels}. For example, the two lowest levels with negative $s$ ($1,-$ and $2,-$)  should cross the  $N_L=0$ (insensitive to the SO interaction) level as depicted in Fig.~\ref{levels}. We can therefore expect that the energy gap (and the corresponding QH state) disappears at the Fermi level at some specific magnetic fields $B_{c1}$ and $B_{c2}$ for the filling factors $\nu=1$ and 2, respectively. This simple one-particle picture can be seriously affected by many-body interaction, at least because the Zeeman splitting is known to be subject to exchange-interaction enhancement~\cite{biasiol04,kalin} at odd filling factors.

In the QH regime, the bulk and the edge spectra are connected because of the specific QH topology~\cite{halperin}. Thus, the bulk spectrum reconstruction should be revealed in the edge spectrum which can be probed in transport experiment~\cite{ESreview}.

Here, we experimentally demonstrate the simultaneous reconstruction of the  bulk and the  edge energy spectra  caused by strong spin-orbit interaction, at the two lowest filling factors $\nu=1$ and $\nu=2$ for $In_xGa_{1-x}As$ two-dimensional electron system with $x = 0.75$. The observed reconstruction is of different character at these filling factors: it is characterized by zeroing of the energy gap at the levels crossing point at $\nu=2$, while the reconstruction at $\nu=1$ goes through the coexistence of two $\nu=1$ quantum Hall phases with different spin projections. An analysis indicates a strong influence of many-body interaction on the reconstruction at $\nu=1$.

\section{Samples and technique}

Our samples are grown by solid source molecular beam epitaxy on semi-insulating GaAs  substrates. The active layer is composed of a 20-nm thick $In_{0.75}Ga_{0.25}As$ quantum well sandwiched between a lower 50-nm thick and an upper 120-nm thick $In_{0.75}Al_{0.25}As$ barriers. Details on the growth parameters can be found elsewhere~\cite{biasiol05}. The samples are not intentionally doped, but a two-dimensional electron gas is formed in the quantum well due to deep donor levels in the InAlAs barriers~\cite{biasiol04a}. The 2DEG mobility at 4K is about $5 \cdot 10^{5}  $cm$^{2}$/Vs  and the carrier density is   $4.1 \cdot 10^{11}  $cm$^{-2}$.

\begin{figure}
\centerline{\includegraphics*[width=\columnwidth]{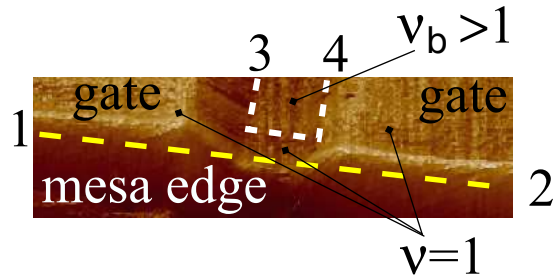}}%
\caption{(Color online) AFM image of the narrow (2~$\mu$m) gate-gap region at the outer mesa edge, see for details Refs.~\protect\onlinecite{ESreview}. For local intra-edge transport investigations, the 2DEG under the gate is kept at the integer value $\nu=1$ by corresponding adjustment of the gate voltage $V_g$. The stripes of compressible electron liquid~\protect\cite{shklovskii} are schematically drawn (thick dash lines) along the gate edges for the  bulk filling factor $\nu_b=2$. They are separated by a $\nu=1$ incompressible stripe in the gate-gap. The compressible stripes are at the electrochemical potentials of the corresponding Ohmic contacts (denoted by numbers). We study electron transport across the edge by applying dc current $I$ between the contacts 1,3 and measuring voltage drop $V$ between the contacts 2,4.
\label{sample}}
\end{figure}

The samples are patterned in a quasi-Corbino sample geometry, see for details, Refs.~\protect\onlinecite{ESreview,alida,zeeman}. The mesa (about 2x2~mm$^2$) has an etched central region ($\sim 0.5\times 0.5\mbox{mm}^2$). Ohmic contacts are placed both at inner and at outer  mesa edges. A metallic top gate covers the area between two mesa edges. Thus, because of macroscopic size, the samples allows standard bulk magnetotransport measurements in Corbino topology.

The possibility of local intra-edge transport investigations is provided by a single gate-free region, which connects two mesa edges. This region is  narrow (2~$\mu$m) at the outer mesa edge, see Fig.~\ref{sample}. At the inner edge, this region is of macroscopic size, so the inner Ohmic contacts are connected to the gate-free 2DEG. 

The details of both experimental techniques are given in Sections~\ref{bulk},\ref{edge} respectively. To avoid a gate leakage at the mesa step~\cite{biasiol08}, a 400~nm SiO isolator layer is thermally evaporated before processing of the gate. Thick  SiO layer   reliably covers the mesa step, so there is no detectable leakage current at low temperatures in the used gate voltage range.  All measurements are performed in a dilution refrigerator at the minimal temperature of 60~mK. The obtained results are independent of the cooling cycle.

\begin{figure}
\includegraphics*[width=\columnwidth]{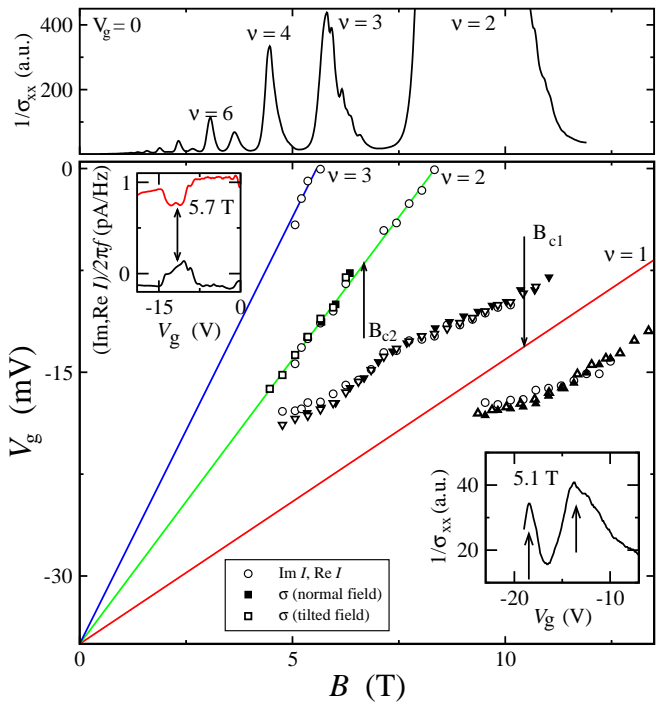}%
\caption{(Color online) (Upper panel) A number of magnetoresistance $1/\sigma_{xx} (B)$ oscillations at zero gate voltage $V_g=0$. (Main figure) Landau fan chart for 2DEG under the gate in $(B,V_g)$ plane. Each point is obtained as the center of the minimum/maximum structure in  the magnetocapasitance signal (open circles, left inset) or the magnetoresistance one (other symbols, right inset). The points, obtained by these two methods coincide well with each other. The magnetoresistance data are demonstrated for normal (filled symbols) and tilted (open symbols, $\Theta=27^\circ$) magnetic fields. They coincide well when plotted against the normal field component, so the fan structure is not sensitive to the in-plane field. The data points for $\nu=2,3$ are concentrated along   straight lines. The lines start from the same onset voltage at zero field, and their slopes relate as the filling factor ratio $3/2$. This behavior confirms the  proportionality of the electron concentration to the gate voltage. There are no points around $B_{c2}=6.7$~T for $\nu=2$, demonstrating the absence of the QH state in this region.  By contrast,  the data points for $\nu=1$   deviate from the expected fan line, despite of the same gate voltage region.  They are concentrated along   two overlapping branches, which indicates the competition between two QH states at $\nu=1$ around the critical field $B_{c1}=10.4$~T.  At lower concentrations, $\nu=1,2$  QH states appear only around $\approx 5$~T, because of much lower mobility.
\label{fan}}
\end{figure}

\section{Experimental results: Landau fan chart}\label{bulk}

We study the bulk spectrum reconstruction qualitatively by drawing the so-called Landau fan chart, see Fig.~\ref{fan}. Each data point indicates the position of the corresponding QH state in $B,V_g$ plane. To obtain these points, we apply two different methods. 

The first one is based on the so-called magnetocapacitance technique~\cite{dqw}. The advantage of this technique is  that the gate-free region has no influence on the magnetocapasitance signal. We modulate the dc gate bias $V_g$ by a small ac voltage (12~mV) and measure an ac current between the gate and one of the Ohmic contacts, see the left inset to Fig.~\ref{fan}. In this case the imaginary part of the current $Im(I)$ reflects the sample capacitance, while the active component $Re(I)$ is proportional to the $1/\sigma_{xx}$ under the gate, see, e.g., Ref.~\onlinecite{dqw} and references therein. 

In the QH regime the bulk of the sample is incompressible, so $Im(I)$ is a minimum. On the other hand,  $\sigma_{xx}$ tends to zero in this regime, which is reflected by a peak in $Re(I)$. Thus, the developed minimum/maximum structure in $Im(I), Re(I)$ indicates a QH regime. We use a center of this minimum/maximum structure as a position of the QH state in $(B,V_{g})$ plane~\cite{highfreq}, see Fig.~\ref{fan}.

It is also possible to trace  $1/\sigma_{xx}$ directly as a two-point resistance between one of the inner and one of the outer Ohmic contacts, see the right inset to Fig.~\ref{fan}. The gate-free region is of minimal influence on the two-point resistance because of negligible charge transfer in the narrow gate-gap region at the outer mesa edge, see the Section~\ref{edge} for details. The geometry is still not a perfectly symmetric Corbino disk, so the  measured signal includes an unknown geometrical factor. We also depict the positions of the measured  $1/\sigma_{xx}$ peaks in the main field of Fig.~\ref{fan}. 

Good coincidence between the data points, obtained by two different methods, guarantees the correctness of the QH state position determination. The methods support each other especially for broad  minimum/maximum structures like the ones presented in the insets to Fig.~\ref{fan}, or, on the contrary, for low signals. 

The data points for $\nu=2,3$ filling factors are concentrated along   straight lines in Fig.~\ref{fan}.  These lines start from the same onset voltage at zero field, and their slopes relate as the filling factor ratio $3/2$.  
We reliably determine the filling factor numbers from a wide set of $1/\sigma_{xx} (B)$ oscillations at zero gate voltage, see the upper panel in Fig.~\ref{fan}. This behavior confirms the proportionality of the electron concentration to the gate voltage in our samples. The data points appear only at $\approx 5$~T in low fields in  Fig.~\ref{fan}, because the 2DEG mobility decreases strongly at lower 2DEG concentrations~\cite{biasiol08}.

The experimental fan chart in Fig.~\ref{fan} has three peculiarities. First, it depends only on the normal field component at low tilt angles ($\Theta=0,27^\circ$). This is in  good correspondence with results of Ref.~\onlinecite{biasiol04}, where the influence of the in-plane component was only observed at the highest ($\Theta>80^\circ$) tilt angles. Second, there are no data points around $B_{c2}=6.7$~T for $\nu=2$, demonstrating the  absence of the $\nu=2$ QH state in this region. Third, the data points for $\nu=1$   deviate from the expected fan line. They are concentrated along   two overlapping branches. Since these data are obtained in the same $V_g$ range as for $\nu=2$, the electron density in fact follows the gate bias. We have to conclude, that the two overlapping branches indicate~\cite{vtd06} a competition between two QH states at $\nu=1$ around the critical field $B_{c1}=10.4$~T. 

Landau fan charts only qualitatively reflect the presence of an energy gap at the Fermi level at some $(B,V_g)$. The $In_{0.75}Ga_{0.25}As$ quantum wells are characterized by a long-range In concentration modulation~\cite{biasiol08}. This results in long-range potential fluctuations in the well, which can also be seen~\cite{nichols,aristov} from wide minimum/maximum structures in the insets to Fig.~\ref{fan}. For this reason, standard bulk quantitative measurements of the gap~\cite{nichols,aristov} will  give the gap values averaged over the sample plane.  Instead, we perform {\it local} measurements of the energy gap at the Fermi level at $\nu=1,2$ by means of the edge transport.             

\section{Experimental results: local edge investigations}\label{edge}

\begin{figure}
\includegraphics*[width=\columnwidth]{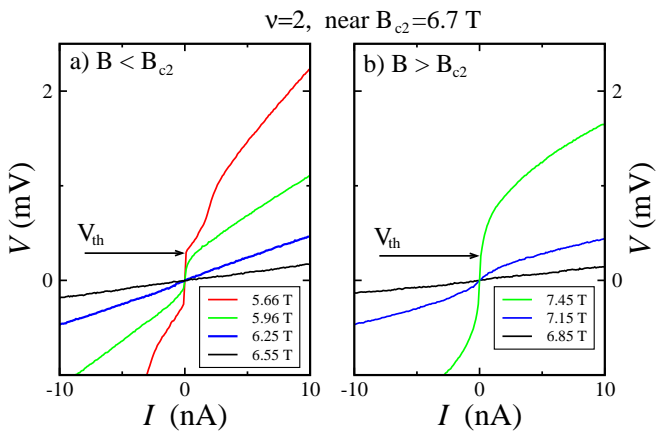}%
\caption{(Color online) $I-V$ curves for transport across $\nu_c=2$ incompressible stripe at the sample edge. The evolution of  the curves is shown along the $\nu=2$ fan line in Fig.~\protect\ref{fan} by approaching $B_{c2}=6.7$~T from low $B<B_{c2}$ (a) or high $B>B_{c2}$ (b) magnetic fields. Far from the critical region, $I-V$ curves are strongly non-linear, the significant current appears only above some threshold voltage $V_{th}$. The threshold value $V_{th}$ characterizes~\cite{ESreview,zeeman} the energy gap $\Delta$ within the $\nu_c=2$ incompressible stripe: $eV_{th}=\Delta$. It tends to zero by approaching $B_{c2}$, which indicates closing of the energy gap at the Fermi level at $B_{c2}$.
\label{IV32}}
\end{figure}

The quasi-Corbino sample geometry is specially designed for local intra-edge transport investigations within the narrow gate-gap region at the mesa edge~\cite{ESreview}. In particular, it allows to obtain the energy gap $\Delta$ within the incompressible stripe at the sample edge from the analysis of $I-V$ curves~\cite{alida,zeeman}.

Since we know the fan chart in Fig.~\ref{fan}, we deplete 2DEG to an integer filling factor $\nu$ at a given magnetic field $B$.  The gate-gap region in Fig.~\ref{sample} is not covered by the gate, so it is characterized by a higher bulk filling factor $\nu_b>\nu$. In this technique, $\nu$ is necessary integer, while $\nu_b$ is not.  For the case of simplicity, the Fig.~\ref{sample} is drawn for both integer filling factors $\nu_b=2$ and $\nu=1$.  

Approaching the mesa edge in the gate-gap, the edge potential diminishes the 2DEG concentration in a so-called compressible region at the mesa edge~\cite{shklovskii}. Since the potential of the etched edge is smooth~\cite{ESreview}, the local filling factor $\nu_c$ is  diminishing smoothly from the bulk $\nu_b$ value to zero at the edge~\cite{shklovskii}. Incompressible stripes of finite width are formed~\cite{shklovskii} along the edge near the integer local filling factors $\nu_c<\nu_b$, e.g. at $\nu_c=2$ and 1 for $\nu_b=3$. In particular, there is an incompressible stripe at $\nu_c=\nu$. Thus, despite the presence of the gate-gap region, an entire incompressible QH state at the filling factor $\nu$ separates the outer (1,2) and the inner (3,4) Ohmic contacts in Fig.~\ref{sample}. 

This $\nu$ QH region is surrounded by two compressible stripes, see Fig.~\ref{sample}. They run along the mesa edge and the gate edge to outer (1,2) and inner (3,4) Ohmic contacts, respectively. We measure $I-V$ characteristics of the gate-gap junction by applying dc current $I$ between the contacts 1,3 and measuring dc voltage drop $V$ between the contacts 2,4 in Fig.~\ref{sample}. Because of the macroscopic gate size in our samples, electron transport is only possible across the $\nu$ incompressible stripe in the gate-gap region~\cite{ESreview,alida,zeeman} at integer $\nu$.

\begin{figure}
\includegraphics*[width=\columnwidth]{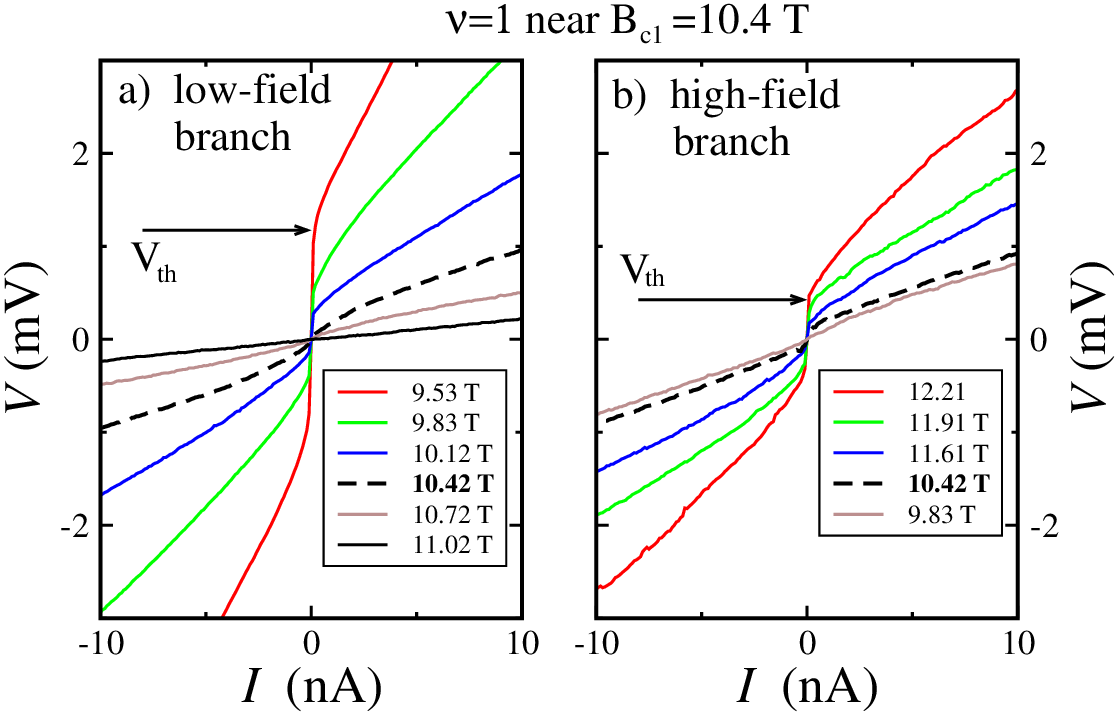}%
\caption{(Color online) $I-V$ curves for transport across $\nu_c=1$ incompressible stripe at the sample edge. The evolution of the curves is demonstrated along the low- (a) or high- (b) field branches of the $\nu=1$ fan line in Fig.~\protect\ref{fan}. Far from $B_{c1}$, $I-V$ curves are also characterized by a strong threshold behavior at low currents. By approaching $B_{c1}$, the $V_{th}$ value is diminishing for low-field branch (a) and disappears together with this branch, cf. Fig.~\protect\ref{fan}. By contrast, $V_{th}$ is monotonously increasing along the high-field branch (b). $I-V$ curves from the different branches coincide around the critical field $B_{c1}=10.4$~T.
\label{IV21}}
\end{figure}

Fig.~\ref{IV32} demonstrates evolution of  experimental $I-V$ curves at different magnetic fields along the $\nu=2$ Landau fan chart line in Fig~\ref{fan}. The experimental curves are strongly nonlinear in the case of developed $\nu=2$ QH state. They tend to be linear in the region of $\nu=2$ QH state disappearance near $B_{c2}=6.7$~T. 

For the filling factor $\nu=1$, $I-V$ curves are presented in Fig.~\ref{IV21}. The $I-V$ non-linearity is monotonically decreasing along the upper (low-field) branch of the fan, see Fig.~\ref{IV21} (a). It is monotonically increasing  along the lower (high-field) branch in Fig.~\ref{IV21} (b).   $I-V$s from two different branches coincide in the magnetic field $B_{c1}=10.4$~T.

The non-linear $I-V$ behavior originates from the edge potential deformation within the $\nu$ incompressible stripe by the applied electrochemical potential imbalance~\cite{alida,zeeman}. In our setup, positive voltage imbalance diminishes the potential jump $\Delta$ within the $\nu$ incompressible stripe. In Figs.~\ref{IV32},\ref{IV21} a noticeable current appears only at some positive threshold voltage $V=V_{th}$, which flattens the edge potential completely~\cite{positive}. Thus, the $V_{th}$ value allows to   estimate~\cite{alida,zeeman,biasiol11} the local energy gap $\Delta=eV_{th}$ within the incompressible stripe at the filling factor $\nu$. 

\begin{figure}
\includegraphics*[width=\columnwidth]{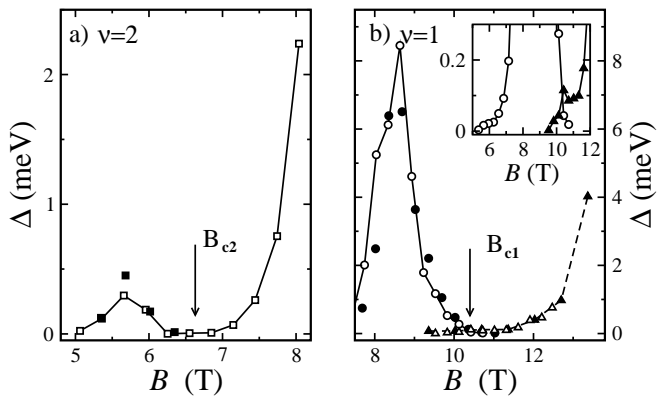}%
\caption{Energy gap $\Delta$ within the $\nu_c=2$ (a) and $\nu_c=1$  (b) incompressible stripes obtained from the threshold $eV_{th}$ of experimental $I-V$ curves $\Delta=eV_{th}$. The data coincide for normal (open symbols) and tilted (filled ones) magnetic field when are plotted against the normal field component $B$. In low fields, non-zero $\Delta$ appears together with the corresponding fan line because of opening the mobility gap at the Fermi level. In higher fields, the gap behavior is non-monotonous. For  $\nu_c=2$ (a) the gap $\Delta$ (squares) is a zero in some region near $B_{c2}=6.7$~T . For  $\nu_c=1$ (b) the gap is diminishing along the upper (low-field) fan branch (circles) and is monotonously increasing along the high-field branch (triangles) . These dependencies intersect each other at the critical field $B_{c1}=10.4$~T. Note, that $\Delta (B)$ is far from the expected linear Zeeman behavior even in the highest accessible fields. 
\label{gap}}
\end{figure}

The obtained $\Delta$ is shown in Fig.~\ref{gap} for both filling factors $\nu=1,2$.  Non-zero gap appears together with the corresponding fan line in Fig.~\ref{fan} and is rising in low fields. At higher fields, the dependence is not monotonous: for the filling factor $\nu=2$ the gap is a zero around $B_{c2}$; for the filling factor $\nu=1$, the gap is diminishing to zero along the upper (low-field) fan branch and is rising along the down (high-field) one. These two dependencies for $\nu=1$ intersect each other in the field $B_{c1}=10.4$~T. Even in the highest accessible fields  $\Delta(B)$ for $\nu=1$ is far from the expected linear Zeeman behavior.

This is exactly the behavior which we can qualitatively expect from the fan chart.  Similarly to the fan chart,  the gap behavior is independent of the in-plane field component at low tilt angles $\Theta=0,27^\circ$. It is important, that the gate-gap region is not covered by the gate. Thus, the observed behavior is determined by 2DEG itself and is not connected with the gate fabrication.

The locally determined gap $\Delta$ is free from the influence of the long-range potential fluctuations in the sample~\cite{zeeman}. These fluctuations can only shift the geometrical position of the continuous $\nu$ incompressible stripe. Thus, $\Delta$ is obtained exactly at the filling factor $\nu$, in contrast to the bulk measurements~\cite{nichols,aristov}. On the other hand, because of the energy levels broadening by the short-range disorder, the gap $\Delta=eV_{th}$ is the mobility gap~\cite{zeeman}, similarly to one obtained in the bulk activation measurements~\cite{nichols}. It is systematically less~\cite{alida,zeeman,biasiol11} than the spectrum gap, also due to electron heating~\cite{biasiol11}. Since the gap is determined by charge transport, it is characterized by $k=\infty$ momentum and is therefore subject to exchange enhancement~\cite{kalin}.

\section{Discussion}

Let us summarize the main experimental results: (i) the simultaneous bulk and edge energy spectra reconstruction occurs at some specific magnetic fields (ii) for $\nu=2$, the gap at the Fermi level $\Delta$ is a zero around $B_{c2}$, and opens to both sides of this region; (iii) for $\nu=1$, there is a competition of different $\nu=1$ QH states around $B_{c1}$ field;  (iv) the gap at $\nu=1,2$ is only controlled by the normal field component for low tilt angles.

This behavior is impossible for a conventional single-subband 2DEG with weak SO coupling. We have to conclude that it originates from the strong spin-orbit coupling in our samples. 

In this case $\nu=2$ QH state behaves as it can be expected from the single-particle Eq.~(\ref{rashba}), see Fig.~\ref{levels}. Because of levels crossing, the energy gap at the Fermi level $\Delta$ is a zero around $B_{c2}$ which results in the disappearance of the $\nu=2$ QH state in this region. 

For $\nu=1$, the coexistence of two QH states around $B_{c1}$ can also be anticipated even from the single-particle Eq.~(\ref{rashba}). As it can be seen from Fig.~\ref{levels}, the transition at $B_{c1}$ is the transition between the two $\nu=1$ spin-polarized QH states with opposite spin polarizations. It is not too astonishing that the transition goes through the coexistence of these phases around $B_{c1}$~\cite{vtd06}. By contrast, two energy levels are filled at $\nu=2$, so the QH state is spin unpolarized. 

The edge transport provides an additional proof for the proposed picture. The non-linear $I-V$ curve indicates the presence of the incompressible stripe in the gate-gap region~\cite{ESreview}. Since $I-V$s are non-linear in Fig.~\ref{IV21} for both branches of the $\nu=1$ fan line, there are two incompressible stripes in the gate-gap region of the sample around $B_{c1}$. Both these stripes belong to the $\nu=1$ QH state but they are deviated from the exact position at the local filling factor $\nu_c=1$, similarly to the $\nu=1$ fan branches. It's worth noting, that the strictly linear $\nu=2$ fan line in Fig.~\ref{fan} ensures the 2DEG macroscopic homogeneity.

An accurate interpretation of our experimental results should  undoubtedly include  many-body effects, which are of primary importance at least at $\nu=1$, see Refs.~\onlinecite{kalin,nichols,aristov}. We are unaware of such theories. The first and very naive attempt is to obtain $B_{c1}$ and $B_{c2}$ from single-particle Eq.~(\ref{rashba}) and to  assume an exchange-enhanced value of $g-$factor for $B_{c1}$ only. The values of the critical fields can be easily calculated from Eq.~(\ref{rashba}): $B_{c1}=\gamma /(2|g|\frac{m}{m_0})$ and $B_{c2}=\gamma /(4+2|g|\frac{m}{m_0})$. Their comparison results in $g^*=2\frac{m_0}{m}(B_{c2}/B_{c1})\approx 30$ at $\nu=1$. This value is not far from that expected for InGaAs systems~\cite{kita,gui,biasiol04}. Moreover, the crucial role  of the exchange enhancement is also confirmed by the the independence of the experimental data of the in-plane field component~\cite{kalin,shmeller,khrapai}. The corresponding SO parameter $\gamma=2|g^*|\frac{m}{m_0}B_{c1}\approx 28$~T is high enough to explain why the gap $\Delta (B)$ for $\nu=1$ is far from the expected linear Zeeman behavior even in the highest fields in Fig.~\ref{gap}. The resulting SO parameter 
$\alpha=\left (\frac{\hbar^3e\gamma}{8m^2}\right )^{1/2}\approx 10^{-10}$~eVm is near the highest values, known for InGaAs systems. 

It's worth noting, that the Dresselhaus-type SO coupling gives the energy spectrum similar~\cite{winkler} to Eq.~(\ref{rashba}). However, the obtained  parameter $\gamma$ is too big for the known Dresselhaus SO coupling constant (133~eV$\AA^3$ in InAs). This is in a good correspondence with the experimental~\cite{nitta} fact that the SO coupling is of Rashba type for $In_xGa_{1-x}As$ narrow asymmetric quantum wells. We wish to mention here, that   consistent accounting for the many-body interaction can affect these estimates.

\section{Conclusion}

As a result, we experimentally demonstrate the simultaneous reconstruction of the  bulk and the  edge energy spectra  caused by strong spin-orbit interaction, at the two lowest filling factors $\nu=1$ and $\nu=2$ for $In_xGa_{1-x}As$ two-dimensional electron system with $x = 0.75$. The observed reconstruction is of different character at these filling factors: it is characterized by zeroing of the energy gap at the levels crossing point at $\nu=2$, while the reconstruction at $\nu=1$ goes through the coexistence of two $\nu=1$ quantum Hall phases with different spin projections. An analysis indicates a strong influence of many-body interaction on the reconstruction at $\nu=1$.

\acknowledgments

We wish to thank  V.T.~Dolgopolov, V.S.~Khrapai, and D.E.~Feldman for fruitful discussions.
We gratefully acknowledge financial support by the RFBR and RAS.


\begin{thebibliography}{99}

\bibitem{datta} S. Datta and B. Das, Appl. Phys. Lett. 56,  665 (1990).
\bibitem{winkler} R. Winkler, Spin-Orbit Coupling Effects in Two-Dimensional Electron and Hole
Systems, Springer Tracts in Modern Physics 191, (2003).

\bibitem{biasiol08} Daniele Ercolani, Giorgio Biasiol, Emiliano Cancellieri, Marcello Rosini, Carlo Jacoboni, Franco Carillo, Stefan Heun, Lucia Sorba, and Frithjof Nolting, Phys. Rev. B 77, 235307 (2008).
\bibitem{nitta} Nitta, T. Akazaki, H. Takayanagi, and T. Enoki, Phys. Rev. Lett. 78, 1335 (1997).
\bibitem{kita} T. Kita, Y. Sato, S. Gozu, and S. Yamada, Physica B 298, 65 (2001) 
\bibitem{gui} Y. S. Gui, C. M. Hu, Z. H. Chen, G. Z. Zheng, S. L. Guo, J. H. Chu, J. X. Chen, and A. Z. Li, Phys. Rev. B 61, 7237 (2000).
\bibitem{biasiol04} W. Desrat, F. Giazotto, V. Pellegrini, F. Beltram, F. Capotondi, G. Biasiol, L. Sorba, and D. K. Maude,  Phys. Rev. B 69, 245324 (2004).
\bibitem{spain} M. A. Hidalgo, R. Cangas, arXiv:1201.2516
\bibitem{bychkov} Yu.A.~Bychkov and E.I.~Rashba, JETP Lett., 39, 66  (1984); 6039 J. Phys. C 17 (1984).
\bibitem{kalin} T. Ando and Y. Uemura, J. Phys. Soc. Jpn. {\bf 37}, 1044 (1974) [CAS]; Yu. A. Bychkov, S. V. Iordanskii, and G. M. Eliashberg, JETP Lett. {\bf 33}, 143 (1981); C. Kallin and B. I. Halperin, Phys. Rev. B {\bf 30}, 5655 (1984).
\bibitem{halperin} B. I. Halperin, Phys.\  Rev.\ B  {\bf
25}, 2185 (1982).

\bibitem{biasiol05} F. Capotondi, G. Biasiol, D. Ercolani, V. Grillo, E. Carlino, F. Romanato, and L. Sorba, Thin Solid Films 484, 400 (2005).
\bibitem{biasiol04a} F. Capotondi, G. Biasiol, I. Vobornik, L. Sorba, F. Giazotto, A. Cavallini, and B. Fraboni, J. Vac. Sci. Technol. B 22, 702 (2004).

\bibitem{ESreview} For a review on intra-edge transport experiments in quasi-Corbino geometry see E. V. Deviatov, Physics-Uspekhi 50 (2) 197 (2007); E. V. Deviatov and A Lorke, phys. stat. sol. (b) 245, 366 (2008).
\bibitem{alida} A. W\"urtz, R. Wildfeuer, A. Lorke, E. V. Deviatov, and V. T. Dolgopolov, Phys. Rev. B {\bf 65}, 075303 (2002).
\bibitem{zeeman} E.V. Deviatov, A. Lorke, G. Biasiol, L. Sorba, W. Wegscheider, JETP Lett. {\bf 92},  69 (2010).


\bibitem{dqw} V. T. Dolgopolov, A. A. Shashkin, E. V. Deviatov,  F. Hastreiter, M. Hartung, A. Wixforth, K. L. Campman, and A. C. Gossard Phys. Rev. B 59, 13235 (1999).
\bibitem{highfreq}  It's worth noting, that the experimental traces like in the left inset to  Fig.~\protect\ref{fan} are obtained in a so-called high frequency regime~\protect\cite{dqw} despite the low ac frequency of 2~Hz.
\bibitem{vtd06} V. T. Dolgopolov, E. V. Deviatov, V. S. Khrapai, D. Reuter, A. D. Wieck, A. Wixforth, K. L. Campman, and A. C. Gossard
Phys. Stat. Sol. (b) 243, 3648  (2006).
\bibitem{nichols} A. Usher, R. J. Nicholas, J. J. Harris, and C. T. Foxon, Phys. Rev. B {\bf 41}, 1129 (1990).
\bibitem{aristov} V. T. Dolgopolov, A. A. Shashkin, A. V. Aristov, D. Schmerek, W. Hansen, J. P. Kotthaus, and M. Holland, Phys. Rev. Lett. \textbf{79}, 729 (1997)


\bibitem{shklovskii} D. B. Chklovskii, B. I. Shklovskii, and L. I. Glazman, Phys. Rev. B {\bf 46}, 4026 (1992).
\bibitem{positive} We use positive $V_{th}>0$ as a measure of the local gap, because in our setup it is {\em positive} imbalance which flattens the edge potential. This was demonstrated experimentally in Refs.~\protect\onlinecite{alida,ESreview,zeeman} from the analysis of strongly asymmetric $I-V$ curves. This asymmetry was a result of strongly different Zeeman and cyclotron gap values in GaAs. This is not the case for the present InGaAs system especially near $B_{c1}$ and $B_{c2}$, see Figs.~\protect\ref{IV32},\protect\ref{IV21}. 
 
\bibitem{biasiol11} Nicola Paradiso, Stefan Heun, Stefano Roddaro, Lucia Sorba, Fabio Beltram, and Giorgio Biasiol, Phys. Rev B 84, 235318 (2011).

\bibitem{shmeller} A. Schmeller, J. P. Eisenstein, L. N. Pfeiffer, and K. W. West, Phys. Rev. Lett. \textbf{75}, 4290 (1995).
\bibitem{khrapai} V. S. Khrapai, A. A. Shashkin, E. L. Shangina, V. Pellegrini, F. Beltram, G. Biasiol, and L. Sorba  Phys. Rev. B {\bf 72}, 035344 (2005).





\end{thebibliography}
\end{document}